\documentclass[conference]{IEEEtran}
\IEEEoverridecommandlockouts
\usepackage{cite,hyperref}
\usepackage{amsmath,amssymb,amsfonts}
\usepackage{graphicx,url}
\usepackage{textcomp}
\usepackage{xcolor}
\usepackage{comment}
\usepackage{amsmath,amssymb} % define this before the line numbering.
\usepackage{color}
\usepackage[noend]{algpseudocode}
\usepackage{algorithmicx,algorithm}
\usepackage{multirow}
\usepackage{subfigure}

\usepackage{bm}

\algblock{Input}{EndInput}
\algnotext{EndInput}
\algblock{Output}{EndOutput}
\algnotext{EndOutput}

\def\BibTeX{{\rm B\kern-.05em{\sc i\kern-.025em b}\kern-.08em
    T\kern-.1667em\lower.7ex\hbox{E}\kern-.125emX}}

\begin{document}

\title{Decomposing User-APP Graph into  Subgraphs for Effective APP and  User Embedding Learning
}
\author{\vspace{0in}\\
\IEEEauthorblockN{Tan Yu, Jun Zhi, Yufei Zhang, Jian Li, Hongliang Fei, Ping Li} \\
\IEEEauthorblockA{Cognitive Computing Lab, Baidu Research}
\IEEEauthorblockA{Baidu Search Ads (Phoenix Nest), Baidu Inc.}
10900 NE 8th St. Bellevue, Washington 98004, USA\\
No. 10 Xibeiwang East Road, Beijing 100193, China \\\\
\{tanyu01, zhijun, zhangyufei,  lijian26, hongliangfei, liping11\}@baidu.com
}

\maketitle

\begin{abstract}
APP-installation information is helpful to describe the user's characteristics. The users with similar APPs installed might share several common interests and behave similarly in some scenarios.
In this work, we learn a user embedding vector based on each user's APP-installation information. Since the user APP-installation embedding is learnable without dependency on the historical intra-APP behavioral data of the user, it complements the intra-APP embedding learned within each specific APP. Thus, they considerably help improve the effectiveness of the personalized advertising in each APP, and they are particularly beneficial for the cold start of the new users in the APP. In this paper, we formulate the APP-installation user embedding learning into a bipartite graph embedding problem. The main challenge in learning an effective APP-installation user embedding is the imbalanced data distribution. In this case,  graph learning tends to be dominated by the popular APPs, which billions of users have installed. In other words, some niche/specialized APPs might have a  marginal influence on  graph learning. To effectively exploit the valuable information from the niche APPs, we decompose the APP-installation graph into a set of subgraphs. Each subgraph contains only one APP node and the users who install the APP. For each mini-batch, we only sample the users from the same subgraph in the training process. Thus, each APP can be involved in the training process in a more balanced manner. After integrating the learned APP-installation user embedding into our online personal advertising platform, we obtained a considerable boost in CTR, CVR, and revenue.\\
\end{abstract}


\section{Introduction}

For different users, a personalized advertising system feeds different ads based on the estimated relevance between the ad and the user's interest. Normally, the relevance between a user and an ad is measured by the similarity between their embeddings, which are learned jointly from the users' historical behaviors on the ads. Nevertheless, for new users, there are no historical user-ad behaviors for learning effective user embedding. This issue of modeling new users is normally defined as the cold start problem. To solve the cold start problem, we usually exploit the user's demographic attributes, such as age, region, and gender. The attribute embedding has been effectively learned based on the ordinary users' rich experience accumulated in the past and can readily generalize well to the new users. Since the attribute embedding does not rely on the historical user behaviors, they are useful for tackling the cold start problem.

This work explores a new type of attribute embedding learned from the APP-installation information. The users who install the same APP might share some common interests and tend to behave similarly. Meanwhile, a user's installed APP lists might encode much richer fine-grained information about a user than basic demographic information, such as age, gender, and location. Thus, if exploiting the users' APP-installation information effectively, we might significantly boost the performance of our personalized advertising platform for the new users. In fact, the APP-installation information benefits not only  the new users but also the regular users. This is because the learned user's APP-installation embedding complements the user's behavior embedding. Thus, when incorporating the APP-installation embedding into our model, we also observed improvements for the regular users' personalized advertising performance.

\begin{figure}

\vspace{0.3in}

\centering
\includegraphics[width=3in]{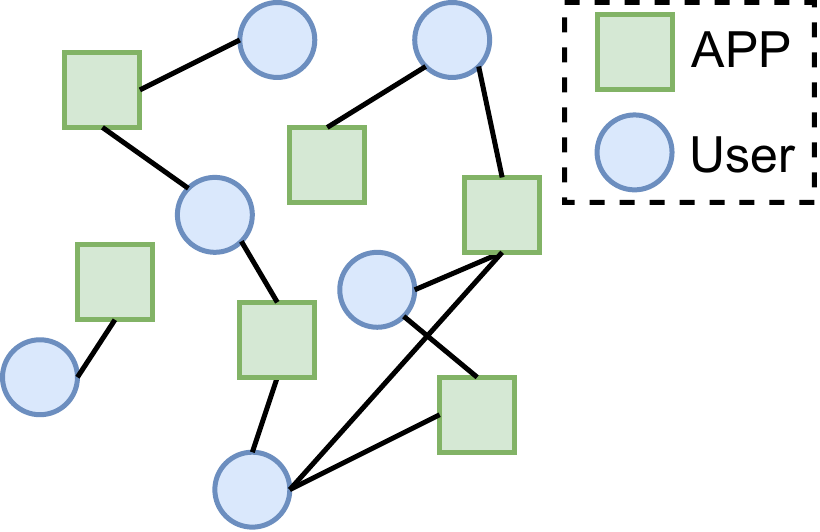}
\caption{The visualization of a user-APP undirected bipartite graph. A node with a blue dot denotes a user and a node with a green box denotes an APP. An edge exists between an APP and a user if the user installs the APP.}
\label{fig:user_app_graph}
\end{figure}

We formulate the APP-installation embedding as a bipartite graph embedding problem. The bipartite graph consists of two types of nodes, including the user nodes and the APP nodes as visualized in Figure~\ref{fig:user_app_graph}. An edge exists between a user node and an APP node if the user has installed the APP in his/her mobile phone. Straightforwardly, we could utilize any existing graph learning methods such as graph convolutional neural network (GCN) to learn the user node embedding and the APP node embedding. Nevertheless, a serious issue caused by imbalanced data distribution makes the training of the graph learning model extremely challenging. Specifically, for a popular APP\footnote{\url{https://www.businessofapps.com/data/most-popular-apps/}}, it is installed by billions of users, generating billions of edges in the graph. In contrast, a niche APP installed by millions of users can only create millions of edges in the graph. In this case, graph learning is dominated by the billions of edges created by the popular APPs, and the edges from the niche APPs might be swamped. But the edges from the popular APPs might not encode useful discriminating information since everyone almost installs them. In contrast, the edges from the niche APPs might be very useful for describing a user's characteristics, but that useful information might not gain enough attention when training the graph embedding.

In this work, we propose a novel sampling approach to tackle the imbalanced data distribution issue for learning effective APP-installation user embeddings. Specifically, we decompose the user-APP graph into a set of sub-graphs. Each subgraph contains only a single APP and the users who install the APP. In the training process, we sample a subgraph for each iteration and construct training triplets based on users within the subgraph for embedding learning. In this manner, the popular APPs and the niche APPs will be   involved in the training process in a fair manner. The offline and online experiments  demonstrate the excellence of our method.

\section{Related Work}
%In this work, we formulate the user embedding learning based on APP-installation information into a graph embedding problem. We review some related works on graph embedding in this section.
% for better adversing .  

%\vspace{0.1in}
\vspace{0.1in}
\noindent \textbf{Factorization based methods}.\;
Factorization-based methods rely on an affinity matrix encoding the connections between nodes in the graph. They factorize the affinity matrix to obtain the  embedding vectors for nodes.  A pioneering work, Laplacian Eigenmaps~\cite{belkin2001laplacian} aims to keep the embedding of two nodes close when the weight of the edge connecting these two nodes is high. It seeks to minimize the weighted summation of squares of distance between nodes  while the weight of each item in the summation corresponds to the weight of the edge. It is formulated into an eigenproblem.  Nevertheless, it is extremely slow when solving the eigen problem in the scenario when the number of nodes is huge. Ahmed \emph{et al.}~\cite{ahmed2013distributed} propose  a framework for large-scale graph decomposition. They  partition a graph based on minimizing the number of neighboring vertices. GraRep~\cite{cao2015grarep} integrates global structural information of the graph into the graph learning process. HOPE~\cite{ou2016asymmetric}  learns the graph embedding for nodes with the asymmetric transitivity,  which is a critical property of the directed graph. \cite{shrivastava2014new} develop a graph kernel methods based on the iterations of power method applied on the adjacency matrix. 

\vspace{0.1in}
\noindent \textbf{Random walks based methods}.\; Random walks are very useful when we have only access to a part of the graph or the graph is too large to be modeled globally.  DeepWalk~\cite{perozzi2014deepwalk} creates multiple random walks, and maximizes the sum of log-likelihoods for each random walk. It preserves higher-order proximity between nodes in the graph.  \emph{node2vec}~\cite{grover2016node2vec} also encodes higher-order proximity between nodes by maximizing the probability of occurrence of subsequent nodes.  It conducts a trade-off between breadth-first searches (BFS) and depth-first searches (DFS) on the graph to generate a more effective graph embedding than DeepWalk.  Walklets~\cite{perozzi2016walklets} additionally incorporates  explicit modeling in random walks. Hierarchical Representation Learning for Networks (HARP)~\cite{chen2018harp} proposes a better initialization strategy  to avoid the local optima in optimization. %Random walk based methods not only can be used to compute a rich/informative representation of the nodes, but also it was shown that they can boost the performance of the graph neural networks~\cite{rahmani2010necessity,rahmani2021non}.

\vspace{0.1in}
\noindent \textbf{Neural network based methods}.\;  SDNE~\cite{wang2016structural} stacks multiple layers of non-linear functions to preserve highly non-linear network structure. It adopts an auto-encoder structure which uses the embedding to reconstruct its neighbors. DNGR~\cite{cao2016deep} feeds  the  positive point-wise mutual information matrix into a stacked denoising autoencoder to  capture higher-order proximity in the learned graph embedding.   Nevertheless, SDNE and DNGR consider the whole graph and take as input the global neighborhood of each node, which are not efficient for large-scale graphs. Recently, graph convolution neural network (GCN) provides an effective and efficient solution by adopting a configuration with local constraints. These methods can be categorizes into spatial-based methods~\cite{micheli2009neural,atwood2016diffusion,hamilton2017inductive,velivckovic2018graph,chiang2019cluster,xu2018powerful} and  spectral-based methods~\cite{bruna2014spectral,henaff2015deep,defferrard2016convolutional,kipf2017semi,levie2018cayleynets,li2018adaptive}.  Spatial-based methods directly  conduct  convolution on the original graph. In contrast, spectral-based  methods conduct convolution on the spectrum of the  adjacent matrix of the graph. In~\cite{rahmani2010necessity,rahmani2021non}, it was shown that the incorporation of the node representation vectors computed by a random walk based method in GCN can effectively boost the performance of the GCN.  

\vspace{0.05in}
\section{Method}
\vspace{0.05in}

In this section, we introduce  graph-based embedding learning for modeling the APP-installation information of users.

\vspace{0.05in}
\subsection{Graph Decomposition}

\begin{figure}[htpb!]

\vspace{0.2in}
\centering
\includegraphics[width=3.7in]{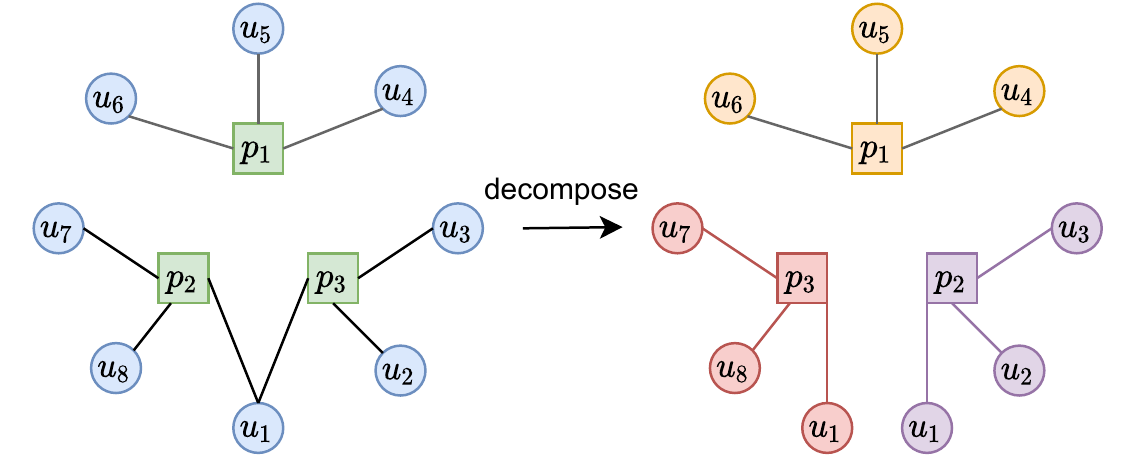}
\caption{The visualization of  decomposing a user-APP graph into subgraphs. The  user-APP graph $\mathcal{G}$ consists of the node set $\mathcal{V} = \{p_i\}_{i=1}^3 \cup \{u_i\}_{i=1}^8 $ where $p_i$ denotes an APP node and $u_i$ is a user node. $\mathcal{G}$ is decomposed into three subgraphs $\{\mathcal{G}_i\}_{i=1}^3$.  $\mathcal{G}_i$ consists of the node set $\mathcal{V}_i$. In this example, $\mathcal{V}_1  =  \{p_1, u_4,u_5,u_6\}$, $\mathcal{V}_2 = \{p_2, u_1,u_2,u_3\}$, and  $\mathcal{V}_3 = \{p_3, u_7,u_8,u_1\}$. Note that, the user $u_1$ is connected with two APPs $p_2$ and $p_3$. Thus, $u_1$ is included in two subgraphs, $\mathcal{G}_1$ and $\mathcal{G}_2$.}
\label{decompose}
\end{figure}

\noindent \textbf{Definition.}\; We denote the set of APPs used for building the graph by $\{{s}_i\}_{i=1}^N$, and denote the set of users  by $\{{t}_j\}_{j=1}^M$.  They constitute the node set $\mathcal{V} = \{{p}_1,\cdots,{p}_N,{u}_1,\cdots,{u}_M\}$. Meanwhile, the edge set $\mathcal{E}$ contains all edges connecting two nodes $\{e_{s_i,t_i}\}_{i=1}^L$, where  $s_i$ denotes the index of the user  and $t_i$ denotes the index of the user in the  $i$-th edge, $e_{s_i,t_i}$. That is,  the existence of edge $e_{s_i,t_i}$ means that the user $u_{s_i}$ has installed the APP  $a_{t_i}$. The user-APP graph $\mathcal{G}$  is constructed based on the node set and the edge set $\{\mathcal{E}, \mathcal{V} \}$. We further define the subgraph $\mathcal{G}_i$ with the node set $\mathcal{V}_i$ and the edge set $\mathcal{E}_i$. $\mathcal{V}_i$ contains only one APP node $p_i$ and the user nodes  connected to $p_i$. $\mathcal{E}_i$ contains all edges which connect the APP node $p_i$. We visualize the process of decomposing a graph into a set of subgraphs in Figure~\ref{decompose}.

\subsection{Graph Learning}
\vspace{0.05in}

\noindent\textbf{Initialization.}  We denote the embedding of the user $u_i$ by $\mathbf{u}_i$ and the embedding of an APP $p_j$ by $\mathbf{p}_j$. We denote the indices of users installing the APP $u_i$ by $\mathcal{I}_i$.
 The user embeddings are randomly initialized. In parallel, an APP embedding $\mathbf{p}_j$ is initialized by averaging the embeddings of users installing the APP:
 \begin{equation}
     \mathbf{p}_j = \frac{\sum_{k \in \mathcal{P}_i} \mathbf{u}_k }{|\mathcal{P}_i|},
 \end{equation}
 where $|\mathcal{P}_i|$ denotes the cardinality of the set $\mathcal{P}_i$, \emph{i.e.}, the number of users installing the APP $p_i$.

\vspace{0.1in}
\noindent \textbf{Subgraph sampling.}
As we mentioned, for a subgraph $\mathcal{G}_i$, it contains an APP node  ($p_i$) and  the users installing the APP $p_i$.  Let us denote the probability of sampling the subgraph $\mathcal{G}_i$ as $P(\mathcal{G}_i)$. A native sampling approach is sampling the sub-graph with a probability proportional to the number of user nodes in the subgraph. That is,
\begin{equation}
\label{naive}
    P(\mathcal{G}_i)  = \frac{|\mathcal{P}_i|}{\sum_{j=1}^N |\mathcal{P}_j|}, \forall i \in [1,N],
\end{equation}
where $N$ is  total number of APPs and $|\mathcal{P}_i|$ denotes the number of users in the subgraph $\mathcal{G}_i$. In this case, each  edge connecting a user and an APP will be involved in the training process with an equal probability. Nevertheless,  this strategy will make the embedding learning dominated by the popular APPs with a huge number of users and the contributions from some niche APPs with a small number of users will be underestimated. To make the contributions from different APPs  balanced, we 
can devise that the sampling probability of each sub-graph to be equal. That is,

\begin{equation}
\label{balance}
    P(\mathcal{G}_i)  = \frac{1}{N}, \forall i \in [1,N].
\end{equation}
In this case, the edges based on niche APPs with a small number of users will be over-sampled, and the edges based on the popular APPs with a huge number of users will be under-sampled. Nevertheless, it might lead to repeatedly sampling for edges from niche APPs, and some edges from the popular APPs might have little chance to be involved in the training process. It tends to make the learned embedding prone to over-fitting due to a lack of diversity in the training samples.  To achieve a balanced sampling and meanwhile suppress over-fitting, we adopt a  trade-off sampling approach. It  devises the probability as 

\begin{equation}
    P(\mathcal{G}_i)  = \frac{|\mathcal{P}_i|^{\tau}}{\sum_{j=1}^N |\mathcal{P}_j|^{\tau}}, \forall i \in [1,N],
\end{equation}
where $\tau$ is a pre-defined positive constant.  Normally, we set $0<\tau<1$. It assigns a higher sampling probability to the subgraph containing more nodes for suppressing over-fitting and meanwhile achieving a good balance among different APPs. When $\tau = 1$, it degenerates to the naive sampling approach defined in Eq.~(\ref{naive}). On the other hand, when $\tau = 0$, it degenerates to the balanced sampling approach defined in Eq.~(\ref{balance}). By default, we set $\tau = 0.5$ in our experiments.

\begin{figure}[htpb!]
\centering
\includegraphics[width=3in]{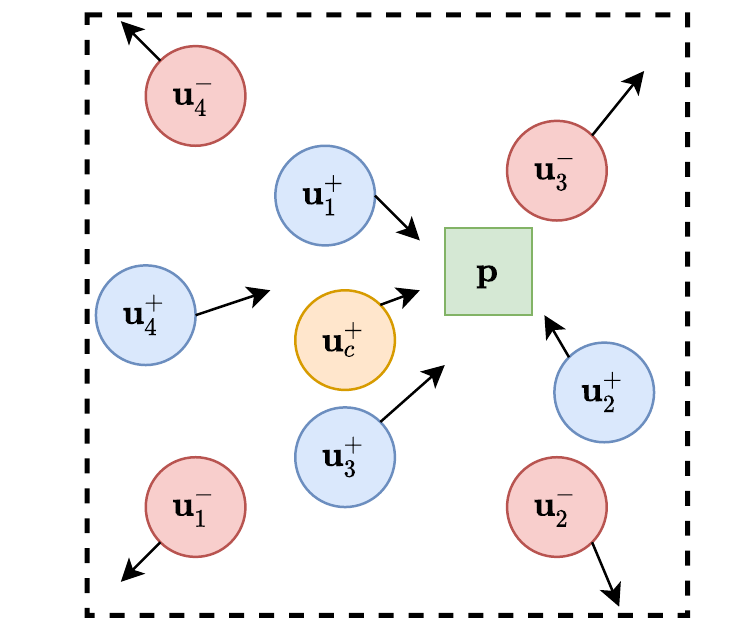}
\caption{The visualization of  user and APP embedding learning within a subgraph. In this example, the APP embedding is $\mathbf{p}$ (green box).  There are four users installing the APP, $\{\mathbf{u}^{+}_i\}_{i=1}^4$ (blue dots) with the centroid $\mathbf{u}^{c}$ (yellow dot) and  four users not installing the APP, $\{\mathbf{u}^{-}_i\}_{i=1}^4$ (red dots). }
\label{decompose2}
\end{figure}

\vspace{0.1in}
\noindent \textbf{Embedding learning within a subgraph.} Let denote the app embedding with a subgraph by $\mathbf{p}$,  the embedding of a user installing the APP by $\mathbf{u}^{+}_i$ and that of a user who does not install the APP by $\mathbf{u}^{-}_j$.    The user and APP embedding learning seeks to keep a large similarity between $\mathbf{p}$ and $\mathbf{u}^{+}_i$. At the same time, it seeks to  maintain a small similarity between  $\mathbf{p}$ and $\mathbf{u}^{-}_j$. Straightforwardly, we can learn the user and the APP embedding through a pairwise loss:

\begin{equation}
    \mathcal{L}_{\mathrm{pair}}  = \frac{1}{n^{+}} \sum_{i=1}^{n^+} \mathrm{log}(1+e^{-\beta s(\mathbf{p}, \mathbf{u}^+_i)}) - \frac{1}{n^{-}} \sum_{i=1}^{n^-} \mathrm{log}(1+e^{-\beta s(\mathbf{p}, \mathbf{u}^-_i)}),
\end{equation}
where $n_{+}$ denotes the number of users installing the APP and $n_{-}$ denotes the number of users who do not install,  $\beta$ is a pre-defined positive constant controlling the softness, and $s(\cdot,\cdot)$ measures the cosine similarity between two vectors, as visualized in Figure~\ref{decompose2}. 

In parallel to the pairwise loss defined above, we devise an additional centroid loss to further enhance the effectiveness of the learned embedding. To be specific, we first compute the centroid of the embeddings of users installing the APP:
\begin{equation}
    \mathbf{u}_c = \frac{1}{n_+} \sum_{i=1}^{n^+} \mathbf{u}^+_i.
\end{equation}
Then the centroid loss is computed by
\begin{equation}
    \mathcal{L}_{\mathrm{centroid}}  =  \mathrm{log}(1+e^{-\beta s(\mathbf{p}, \mathbf{u}^+_c)}).
\end{equation}

\newpage

To stabilize the training, we update the user embedding and the APP embedding in an alternating manner:
\begin{enumerate}
    \item[1)] Fix user embedding $\{\mathbf{u}^+_i\}_{i=1}^{N_+}$ and $\{\mathbf{u}^-_i\}_{i=1}^{N_-}$, and update the APP embedding $\mathbf{p}$ using the centroid loss $\mathcal{L}_{\mathrm{centroid}}$.
    \item[2)] Fix the APP embedding $\mathbf{p}$, and update the positive user embedding $\{\mathbf{u}^+_i\}_{i=1}^{N_+}$ using the  pairwise loss $\mathcal{L}_{\mathrm{pair}}$.
\end{enumerate}
To improve the training efficiency, we achieve this iterative training manner in a parallel way by utilizing the stop-gradient trick. That is, we devise the final loss $\mathcal{L} = \mathcal{L}_{\mathrm{centroid}}+  \mathcal{L}_{\mathrm{pair}}$.
In the meanwhile,  we stop the gradient derived by $\mathcal{L}_{\mathrm{centroid}}$ back-propagating to $\{\mathbf{u}^+_i\}_{i=1}^{N_+}$ and $\{\mathbf{u}^-_i\}_{i=1}^{n_-}$ and meanwhile stop the gradient from $\mathcal{L}_{\mathrm{pair}}$ back-propagating to $\mathbf{p}$ and $\{\mathbf{u}^-_i\}_{i=1}^{n_-}$.

%and denotes the embedding of users installing the APP by $\mathbf{}$

%It tends to lead to the learned embedding prone to over-fitting since 

%It 

%Then we concatenate $\hat{\mathbf{u}}$ with the side-info feature of the A

% Let us denote a positive pair by $\mathcal{P}_+ = <p_+, u_+>$ meaning that the user $u_+$ has installed the APP $p_+$. In parallel, we denote a negative pair by   $\mathcal{P}_- = <p_-, u_->$.

%The user-APP graph is constructed as $\mathcal{G}(\mathcal{V},\mathcal{E})$ where $\mathcal{V}$ contains the nodes 

%\subsection{Multimedia in Advertisement}
%

\vspace{0.2in}
\section{Experiments}

\vspace{0.1in}

\noindent\textbf{Dataset.} To train the model, we collect the information of 80 million users and 50 thousand APPs. On average, each user installs around $30$ APPs.

%We build a graph consisting of $0.6$ billion user nodes and $34$ thousand APPs. Meanwhile, it contains $12$ billion edges connecting user nodes and APP nodes.

%We use each user as the query to retrieve its relevant APPs through 

\vspace{0.1in}
\subsection{Offline experiments}

%To demonstrate the effectiveness of the learned user and APP embedding, we evaluate their memorization capability and  generalization capability.
%The memory performance is measured by the user-APP recordings in our training data, which shows the fitting capability of the user and APP embeddings. In contrast, the inference capability is evaluated by the testing data, which has never been seen in the training process. It demonstrates the generalization capability of the user and APP embeddings.

\vspace{0.1in}
 \noindent \textbf{Memory.} 
 For each APP, we randomly sample $96$ users who have already installed the APP and $96$ users not installing the APP. Note that these APP installation has been involved in the training process. 
 For each APP-user pair, we compute the cosine similarity between their embeddings. Then we threshold the cosine similarity to $0$ or $1$ to predict whether the user has installed the APP or not.
 In Table~\ref{tab:mem}, we show the experimental result. As shown in the table, in the training data, the learned embedding can achieve a $0.953$ precision and $0.981$ AUC, which demonstrates the powerful fitting capability of the learned embeddings.

 %In Table~\ref{} 

 \begin{table}[htpb!]
 
     \centering
      \caption{The memory performance of the learned user and APP embeddings. We report the prediction precision and area-under-curve (AUC) for the APP installation.  }
     \begin{tabular}{c|c}\hline
       Precision   & 0.953  \\\hline
       AUC      & 0.981 \\ \hline
     \end{tabular}
    
     \label{tab:mem}
 \end{table}

 % an installation prediction experiment. This experiment consists of two parts. In the first part, 

 \begin{table}[htpb!]
 
     \centering
      \caption{The inference performance of the learned user and APP embeddings. AUC$^{+}$ denotes the AUC excluding APPs with more than $8\%$  users and  AUC$^{*}$ denotes the AUC  excluding APPs with more than $2.5\%$ users. }
     \begin{tabular}{c|c|c|c} \hline
        & AUC & AUC$^{+}$ & AUC$^{*}$  \\\hline
        APP-side   & 0.797 & 0.840 &  0.854 \\\hline
       User-side & 0.786 & 0.829 & 0.844 \\ \hline
     \end{tabular}
    
     \label{tab:inf}
 \end{table}

 \vspace{0.1in} 
 
 \noindent \textbf{Inference.} 
 To evaluate the inference performance of the learned user and APP embedding, we report the  classification AUC  on the user side and that on the APP side. The user-side AUC is averaged over users. For each user, we test the prediction accuracy using several APPs the user has installed and several APPs the user does not install. 
 The APP-side AUC is measured in a similar manner but is averaged over APPs. Note that the testing cases for inference are not involved in the training process. To be specific, our whole data is collected during $N$ days. We use the data in the first $N-5$ days for training and that from the last $5$ days for testing. Meanwhile, we also report the AUC without excluding APPs with a huge number of users. To be specific, we report  AUC$^{*}$,  which excludes APPs with more than $2.5\%$ users. We also report AUC$^{+}$, which excludes that with more than $8\%$  users. As shown in Table~\ref{tab:inf}, the AUC achieved in the inference is lower than that in Table ~\ref{tab:mem}. %This is expected since the samples for testing the memory performance have been optimized in the training phase. 
In the meanwhile, by excluding some APPs with a huge number of users, AUC$^{+}$ and AUC$^{*}$ are larger than  AUC. %This is also expected since it is difficult to model the APPs with a huge number of users due to the imbalanced data distribution.

\vspace{0.1in}
\noindent \textbf{Ablation study} %Recall that we devise a centroid loss function $\mathcal{L}_{\mathrm{centroid}}$ beside the pairwise loss $\mathcal{L}_{\mathrm{pair}}$. Also, we utilize the stop-gradient strategy for stabilizing the training.
Here, we investigate the influence of removing   $\mathcal{L}_{\mathrm{centroid}}$ or the stop-gradient strategy through ablation study.
As shown in Table~\ref{tab:abl}, when removing  $\mathcal{L}_{\mathrm{centroid}}$, the AUC drops from $0.981$ to $0.977$ and the precision decreases from $0.953$ to $0.948$. Meanwhile, without the stop-gradient strategy, both the AUC and the precision decrease considerably.

\begin{table}[htpb!]
 
     \centering
      \caption{The influence of removing   $\mathcal{L}_{\mathrm{centroid}}$ or the stop-gradient strategy in inference. }
     \begin{tabular}{c|c|c|c} \hline
        & Ours & w/o $\mathcal{L}_{\mathrm{centroid}}$  & w/o stop-gradient  \\\hline
        Precision  & 0.953 & 0.948 & 0.942 \\ \hline
        AUC  & 0.981 & 0.977 &  0.973 \\\hline
     \end{tabular}
    
     \label{tab:abl}
 \end{table}

\subsection{Online experiments} 

We have integrated the user embedding learned from the APP-installation information as a feature which complements the existing user embedding learned from historical behaviors. After launching it in our online personalized advertising platform, we achieved a $+1.1\%$ CTR improvement, a $+1.7\%$ CVR boost,  a +$2.6\%$ increase in revenue as shown in Table~\ref{tab:online}.

%We show the influence of integrating the learned user embedding from user-APP graph in our advertising platform. The experiments are conducted from May. 1st, 2021 to May. 7th, 2021. We report its influence on click-through ratio (CTR) and conversion ratio (CVR) in Table~\ref{online}. As shown in the table, both CTR and CVR improve considerably after integrating the learned user embedding from the user-APP graph in our advertising platform. 

\begin{table}[htpb!]
 
     \centering
      \caption{The online experiments in our online advertising platform during one week. }
     \begin{tabular}{c|c|c} \hline
         CTR & CVR  & Revenue  \\\hline
         $+1.1\%$ & $+1.7\%$ & $+2.6\%$ \\\hline
     \end{tabular}
    
     \label{tab:online}
 \end{table}

\section{Conclusion}
In this paper, we exploit the APP-installation information to assist in modeling the user's characteristics for personalized advertising. To this end, we build a user-APP bipartite graph and adopt a graph convolution network to learn the user embedding. 
We use the learned user embedding from our  user-APP graph as the complementary information to the existing user representation learned from the user profile and the user's historical behaviors. After deploying it in our advertising platform, both CTR and CVR improve considerably.

\bibliographystyle{plain}
\bibliography{refs_scholar}

\begin{thebibliography}{10}

\bibitem{ahmed2013distributed}
Amr Ahmed, Nino Shervashidze, Shravan~M. Narayanamurthy, Vanja Josifovski, and
  Alexander~J. Smola.
\newblock Distributed large-scale natural graph factorization.
\newblock In {\em Proceedings of the 22nd International World Wide Web
  Conference (WWW)}, pages 37--48, Rio de Janeiro, Brazil, 2013.

\bibitem{atwood2016diffusion}
James Atwood and Don Towsley.
\newblock Diffusion-convolutional neural networks.
\newblock In {\em Advances in Neural Information Processing Systems (NIPS)},
  pages 1993--2001, Barcelona, Spain, 2016.

\bibitem{belkin2001laplacian}
Mikhail Belkin and Partha Niyogi.
\newblock Laplacian eigenmaps and spectral techniques for embedding and
  clustering.
\newblock In {\em Advances in Neural Information Processing Systems (NIPS)},
  pages 585--591, Vancouver,Canada], 2001.

\bibitem{bruna2014spectral}
Joan Bruna, Wojciech Zaremba, Arthur Szlam, and Yann LeCun.
\newblock Spectral networks and locally connected networks on graphs.
\newblock In {\em Proceedings of the 2nd International Conference on Learning
  Representations (ICLR)}, Banff, Canada, 2014.

\bibitem{cao2015grarep}
Shaosheng Cao, Wei Lu, and Qiongkai Xu.
\newblock {GraRep}: Learning graph representations with global structural
  information.
\newblock In {\em Proceedings of the 24th {ACM} International Conference on
  Information and Knowledge Management (CIKM)}, pages 891--900, Melbourne,
  Australia, 2015.

\bibitem{cao2016deep}
Yue Cao, Mingsheng Long, Jianmin Wang, Qiang Yang, and Philip~S. Yu.
\newblock Deep visual-semantic hashing for cross-modal retrieval.
\newblock In {\em Proceedings of the 22nd {ACM} {SIGKDD} International
  Conference on Knowledge Discovery and Data Mining (KDD)}, pages 1445--1454,
  San Francisco, CA, 2016.

\bibitem{chen2018harp}
Haochen Chen, Bryan Perozzi, Yifan Hu, and Steven Skiena.
\newblock {HARP:} hierarchical representation learning for networks.
\newblock In {\em Proceedings of the Thirty-Second {AAAI} Conference on
  Artificial Intelligence (AAAI)}, pages 2127--2134, New Orleans, LA, 2018.

\bibitem{chiang2019cluster}
Wei{-}Lin Chiang, Xuanqing Liu, Si~Si, Yang Li, Samy Bengio, and Cho{-}Jui
  Hsieh.
\newblock {Cluster-GCN}: An efficient algorithm for training deep and large
  graph convolutional networks.
\newblock In {\em Proceedings of the 25th {ACM} {SIGKDD} International
  Conference on Knowledge Discovery {\&} Data Mining (KDD)}, pages 257--266,
  Anchorage, AK, 2019.

\bibitem{defferrard2016convolutional}
Micha{\"{e}}l Defferrard, Xavier Bresson, and Pierre Vandergheynst.
\newblock Convolutional neural networks on graphs with fast localized spectral
  filtering.
\newblock In {\em Advances in Neural Information Processing Systems (NIPS)},
  pages 3837--3845, Barcelona, Spain, 2016.

\bibitem{grover2016node2vec}
Aditya Grover and Jure Leskovec.
\newblock node2vec: Scalable feature learning for networks.
\newblock In {\em Proceedings of the 22nd {ACM} {SIGKDD} International
  Conference on Knowledge Discovery and Data Mining (KDD)}, pages 855--864, San
  Francisco, CA, 2016.

\bibitem{hamilton2017inductive}
William~L. Hamilton, Zhitao Ying, and Jure Leskovec.
\newblock Inductive representation learning on large graphs.
\newblock In {\em Advances in Neural Information Processing Systems (NIPS)},
  pages 1024--1034, Long Beach, CA, 2017.

\bibitem{henaff2015deep}
Mikael Henaff, Joan Bruna, and Yann LeCun.
\newblock Deep convolutional networks on graph-structured data.
\newblock {\em arXiv preprint arXiv:1506.05163}, 2015.

\bibitem{kipf2017semi}
Thomas~N. Kipf and Max Welling.
\newblock Semi-supervised classification with graph convolutional networks.
\newblock In {\em Proceedings of the 5th International Conference on Learning
  Representations (ICLR)}, Toulon, France, 2017.

\bibitem{levie2018cayleynets}
Ron Levie, Federico Monti, Xavier Bresson, and Michael~M. Bronstein.
\newblock {CayleyNets}: Graph convolutional neural networks with complex
  rational spectral filters.
\newblock {\em {IEEE} Trans. Signal Process.}, 67(1):97--109, 2019.

\bibitem{li2018adaptive}
Ruoyu Li, Sheng Wang, Feiyun Zhu, and Junzhou Huang.
\newblock Adaptive graph convolutional neural networks.
\newblock In {\em Proceedings of the Thirty-Second {AAAI} Conference on
  Artificial Intelligence (AAAI)}, pages 3546--3553, New Orleans, LA, 2018.

\bibitem{micheli2009neural}
Alessio Micheli.
\newblock Neural network for graphs: {A} contextual constructive approach.
\newblock {\em {IEEE} Trans. Neural Networks}, 20(3):498--511, 2009.

\bibitem{ou2016asymmetric}
Mingdong Ou, Peng Cui, Jian Pei, Ziwei Zhang, and Wenwu Zhu.
\newblock Asymmetric transitivity preserving graph embedding.
\newblock In {\em Proceedings of the 22nd {ACM} {SIGKDD} International
  Conference on Knowledge Discovery and Data Mining (KDD)}, pages 1105--1114,
  2016.

\bibitem{perozzi2014deepwalk}
Bryan Perozzi, Rami Al{-}Rfou, and Steven Skiena.
\newblock Deepwalk: online learning of social representations.
\newblock In {\em Proceedings of the 20th {ACM} {SIGKDD} International
  Conference on Knowledge Discovery and Data Mining (KDD)}, pages 701--710,
  2014.

\bibitem{perozzi2016walklets}
Bryan Perozzi, Vivek Kulkarni, and Steven Skiena.
\newblock Walklets: Multiscale graph embeddings for interpretable network
  classification.
\newblock {\em arXiv preprint arXiv:1605.02115}, 2016.

\bibitem{rahmani2010necessity}
Mostafa Rahmani and Ping Li.
\newblock The necessity of geometrical representation for deep graph analysis.
\newblock In {\em Proceedings of the 20th {IEEE} International Conference on
  Data Mining (ICDM)}, pages 1232--1237, 2020.

\bibitem{rahmani2021non}
Mostafa Rahmani, Rasoul Shafipour, and Ping Li.
\newblock Non-local feature aggregation on graphs via latent fixed data
  structures.
\newblock In {\em Proceedings of the 55th Asilomar Conference on Signals,
  Systems, and Computers (Asilomar)}, pages 1551--1557, 2021.

\bibitem{shrivastava2014new}
Anshumali Shrivastava and Ping Li.
\newblock A new space for comparing graphs.
\newblock In {\em Proceedings of the 2014 {IEEE/ACM} International Conference
  on Advances in Social Networks Analysis and Mining (ASONAM)}, pages 62--71,
  2014.

\bibitem{velivckovic2018graph}
Petar Velickovic, Guillem Cucurull, Arantxa Casanova, Adriana Romero, Pietro
  Li{\`{o}}, and Yoshua Bengio.
\newblock Graph attention networks.
\newblock In {\em Proceedings of the 6th International Conference on Learning
  Representations (ICLR)}, 2018.

\bibitem{wang2016structural}
Daixin Wang, Peng Cui, and Wenwu Zhu.
\newblock Structural deep network embedding.
\newblock In {\em Proceedings of the 22nd {ACM} {SIGKDD} International
  Conference on Knowledge Discovery and Data Mining (KDD)}, pages 1225--1234,
  San Francisco, CA, 2016.

\bibitem{xu2018powerful}
Keyulu Xu, Weihua Hu, Jure Leskovec, and Stefanie Jegelka.
\newblock How powerful are graph neural networks?
\newblock In {\em Proceedings of the 7th International Conference on Learning
  Representations (ICLR)}, 2019.

\end{thebibliography}

\end{document}